\documentstyle[aps]{revtex}

\begin{document}

\title{Ground states of the atoms H, He,\ldots, Ne 
and their singly positive ions in strong
magnetic fields: The high field regime}
\author{M. V. Ivanov\dag\  and P. Schmelcher}
\address{Theoretische Chemie, Physikalisch--Chemisches Institut,
Universit\"at Heidelberg, INF 229, D-69120 Heidelberg,
Federal Republic of Germany\\
\dag Permanent address: Institute of Precambrian Geology and Geochronology,
Russian Academy of Sciences,
Nab. Makarova 2, St. Petersburg 199034, Russia
}

\date{\today}
\maketitle

\begin{abstract}
The electronic structure of the ground and some excited states 
of neutral atoms with the nuclear charge numbers $1\leq Z \leq 10$ 
and their single positive ions 
are investigated by means of our 2D mesh Hartree-Fock method 
for strong magnetic fields $0.5\leq \gamma \leq 10000$. 
For $\gamma=10000$ the ground state configurations 
of all the atoms and ions considered are given by 
fully spin-polarized configurations 
of single-electron orbitals with magnetic quantum numbers ranging  
from $m=0$ to $m=-N+1$ where $N$ is the number of 
the electrons. 
Focusing on the fully spin polarized situation we provide 
critical values of the magnetic field strength for 
which crossovers with respect to the spatial symmetries of the ground state
take place.
It is found that the neutral atoms and singly charged positive ions 
with $2\leq Z \leq 5$ have one 
fully spin-polarized ground state configuration 
whereas for $6\leq Z \leq 10$ one intermediate 
fully spin-polarized configuration with an orbital
of $2p_0$ type occurs. 
\end{abstract}


\section{Introduction}
 
The behavior and properties of atoms in strong magnetic fields 
is a subject which has attracted the interest of many researchers.  
Partially this interest is  
motivated by the astrophysical discovery of strong fields
on white dwarfs and neutron stars \cite{NStar1,NStar2,Whdwarf1}. 
On the other hand, the competition of the diamagnetic and Coulomb
interaction, characteristic for atoms in strong magnetic fields, 
causes a rich variety of complex properties which are
of interest on their own.

Investigations on the electronic structure in the presence of
a magnetic field appear to be quite complicated
due to the intricate geometry of these quantum problems.
Most of the investigations in the literature 
focus on the hydrogen atom
(for a list of references see, for example, \cite{RWHR,Fri89,Kra96,Ivanov88}).
These studies provided us with a detailed understanding 
of the electronic structure of the hydrogen atom in magnetic fields 
of arbitrary strengths. 
As a result the absorption features of certain magnetic white dwarfs 
could be explained and this
allowed for a modeling of their atmospheres (see ref.\cite{Rud94} for a
comprehensive review of atoms in strong magnetic fields 
and their astrophysical applications up to 1994 
and ref.\cite{Schm98} for a more recent review on
atoms and molecules in external fields).
On the other hand there are a number of magnetic white dwarfs whose
spectra remain unexplained and cannot be interpreted in terms of
magnetized atomic hydrogen. Furthermore new magnetic objects
are discovered (see, for example, Reimers et al \cite{Reim98}
in the course of the Hamburg ESO survey) whose spectra await to be explained.
Very recently significant progress has been achieved with respect to the
interpretation of the observed spectrum of the prominent white dwarf GD229 
which shows a rich spectrum ranging from the UV to the near IR.
Extensive and precise calculations on the helium atom provided data
for many excited states in a broad range of field strengths \cite{Bec99}. The comparison
of the stationary transitions of the atom with the positions of the absorption edges
of the observed spectrum yielded strong evidence for the existence of helium
in the atmosphere of GD229 \cite{JSBS229}.

For atoms with several electrons there are two decisive factors which 
enrich the possible changes in the electronic structure with varying
field strength compared to the one-electron system. 
First we have a third competing interaction which is
the electron-electron repulsion and second the different electrons
feel very different Coulomb forces, i.e. possess different one particle energies,
and consequently the regime of the intermediate field strengths
appears to be the sum of the intermediate regimes for the separate electrons.

There exist a number of investigations on two-electron atoms in the literature
(see ref. \cite{Bec99} and references therein). 
Focusing on systems with more than two electrons however the number of investigations is very 
scarce \cite{Neuhauser,JonesOrtiz,Muell84,Godefroid,Ivanov98,IvaSchm98,IvaSchm99}. 
Some of them use the adiabatic approximation in order to investigate the very high field regime. 
These works contain a number of important results on the properties 
and structure of several multielectron atoms.  
Being very useful for high fields the adiabatic approach 
does hardly allow to describe the electronic structure with decreasing field strength: 
particularly the core electrons of multi-electron atoms feel 
a strong nuclear attraction which can be dominated by the external field 
only for very high field strengths. 
In view of this there is a need for further quantum mechanical
investigations on multi-electron atoms, 
particularly in the intermediate to high-field regime.

The ground states of atoms in strong magnetic fields 
have different spatial and spin symmetries in the different regions 
of the field strengths. 
We encounter, therefore, a series of changes i.e. crossovers with respect to 
their symmetries with varying 
field stength. 
The simplest case is the helium atom 
which possesses two ground state configurations: 
the singlet zero- and low-field ground state $1s^2$ and 
the fully spin-polarized high-field ground state $1s2p_{-1}$. 
In the Hartree-Fock approximation the transition point 
between these configurations is given by the field strength $\gamma=0.711$. 
(If not indicated otherwise we use in the following atomic units for all quantities. 
In particular, the magnetic field $\gamma =B/B_0$ is measured in units 
$B_0=\hbar c/ea_0^2=2.3505 {\cdot} 10^5{\rm T}=2.3505 {\cdot} 10^9{\rm G}$.) 
In previous works we have investigated  
the series of transitions of the ground state configurations  
for the complete range of field strengths 
for the lithium \cite{IvaSchm98} and carbon \cite{IvaSchm99} atoms as well as 
the ion ${\rm Li^+}$ \cite{IvaSchm98}. 
The evolution and appearence  of these crossovers and the involved configurations  
become more and more 
intricate with increasing number of electrons of the atom. 
Currently the most complicated atomic system with a completely known sequence of 
ground state electronic configurations 
for the whole range of magnetic field strengths 
is the neutral carbon atom \cite{IvaSchm99}. 
Its ground state experiences six crossovers 
involving seven different electronic configurations 
which belong to three groups of different spin projections 
$S_z=-1,-2,-3$ onto the magnetic field. 
This series of ground state configurations was extracted from results of 
numerical calculations for more than twenty electronic configurations selected via a  
detailed analysis on the basis of general energetical arguments.
The picture of these transitions is especially complicated 
at relatively weak and intermediate fields. 
Due to this circumstance the comprehensive investigation 
of the structure of ground states of atoms is a  
complex problem which has to be solved for each atom separately.
On the other hand, the geometry of the atomic wave functions is simplified 
for sufficiently high magnetic fields: 
Beyond some critical field strength the global ground state is given by 
a fully spin polarized configuration. 
This allows us to push the current state of the art and 
to study the ground states of the full series of neutral atoms and 
singly charged positive ions with $Z\leq 10$, 
i.e. the sequence H, He, Li, Be, B, C, N, O, F, Ne,  in the domain of high magnetic fields. 
For the purpose of this investigation we define the high field domain as the one, 
where the ground state electronic configurations are fully spin polarized 
(Fully Spin Polarized (FSP) regime). 
The latter fact supplies an additional advantage for 
calculations performed in the Hartree-Fock approach, because 
our one-determinantal wave functions are 
eigenfunctions of the total spin operator ${\bf S^2}$.  
Starting from the high-field limit we will investigate the electronic structure 
and properties of the 
ground states with decreasing field strength until we reach the first crossover 
to a partially spin polarized (PSP) configuration, 
i.e. we focus on the regime of field strengths for which 
fully spin polarized configurations represent the ground state.

\section{Method}

The numerical approach applied in the present work coincides with that
of our previous investigations \cite{Ivanov98,IvaSchm98,IvaSchm99}. 
Refs. \cite{Ivanov88,Ivanov98,IvaSchm98,Ivanov94} contain
some more details of the mesh techniques.
We solve the electronic Schr\"odinger equation for the atoms in
a magnetic field under the assumption of an infinitely heavy nucleus 
(see below for comments on finite nuclear mass corrections)
in the (unrestricted) Hartree-Fock approximation.
The solution is established in cylindrical coordinates 
$(\rho,\phi,z)$ with the $z$-axis oriented along the magnetic field.
We prescribe to each electron a definite value of the magnetic
quantum number $m_\mu$.
Each one-electron wave function $\Psi_\mu$ depends on the variables
$\phi$ and $(\rho,z)$ as follows
\begin{eqnarray}
\Psi_\mu(\rho,\phi,z)=(2\pi)^{-1/2}e^{-i m_\mu\phi}\psi_\mu(z,\rho)
\label{eq:phiout}
\end{eqnarray}
where $\mu$ indicates the numbering of the electrons.
The resulting partial differential equations for $\psi_\mu(z,\rho)$
and the formulae for the Coulomb and exchange potentials
have been presented in
ref.\cite{Ivanov94}.

The one-particle equations for the wave functions $\psi_\mu(z,\rho)$
are solved
by means of the fully numerical mesh method described in refs.
\cite{Ivanov88,Ivanov94}.
The feature which distinguishes the present calculations from
those described in ref.\cite{Ivanov94} is the method for the calculation
of the Coulomb and exchange integrals. In the present work as well as in
ref.\cite{Ivanov98,IvaSchm98,IvaSchm99}
we obtain these potentials as solutions
of the corresponding Poisson equation.

Our mesh approach is flexible enough to yield precise
results for arbitrary field strengths.
Some minor decrease of the precision
appears in very strong magnetic fields. 
With respect to the electronic configurations possessing high absolute values 
of magnetic quantum numbers of outer electrons 
some minor computational problems arose also at lower field strengths. 
Both these phenomena are due to a big difference with respect to the
binding energies ${\epsilon_B}_\mu$
of one electron wave functions belonging to the same
electronic configuration
\begin{eqnarray}
{\epsilon_B}_\mu=(m_\mu+|m_\mu|+2s_{z\mu}+1)\gamma/2-\epsilon_\mu
\label{eq:ebinone}
\end{eqnarray}
where $\epsilon_\mu$ is the one electron energy
and $s_{z\mu}$ is the spin $z$-projection.
The precision of our results depends, of course, also on the number of the mesh nodes
and can be improved in calculations with denser meshes.
Most of the present calculations 
are carried out on sequences of meshes with the maximal number of nodes being
$65 \times 65$.

\section{Relevant properties in the high field regime}

In this section we provide some qualitative considerations 
on the problem of the ground states of multi-electron atoms 
in the high field limit. 
These considerations present a starting point for the combined 
qualitative and numerical considerations given in the following section. 
At very high field strengths the nuclear attraction energies and 
HF potentials (which determine the motion along the $z$ axis) 
are small compared to the interaction energies 
with the magnetic field
(which determines the motion perpendicular to the magnetic field
and is responsible for the Landau zonal structure of the spectrum).
Thus in the limit ($\gamma \rightarrow \infty $), 
all the one-electron wave functions of the ground state belong 
to the
lowest Landau zones, i.e. $m_\mu\leq 0$
for all the electrons, and the system must be fully spin-polarized,
i.e. $s_{z\mu}= -{1\over2}$.
For the Coulomb central field the one electron levels form
quasi 1D Coulomb series with the binding energy
$E_B={1\over{2n_z^2}}$ for $n_z>0$, 
whereas $E_B(\gamma \rightarrow \infty)\rightarrow \infty$ for $n_z=0$,
where $n_z$ is the number of nodal
surfaces of the wave function crossing the $z$ axis. 
In the limit $\gamma \rightarrow \infty$ the ground state wave function 
must be formed of the tightly bound single-electron functions with $n_z=0$. 
The binding energies of these functions 
decrease as $|m|$ increases and, 
thus, the electrons must occupy orbitals with increasing $|m|$ starting 
with $m=0$.

In the language of the Hartree-Fock approximation 
the ground state wave function of an atom in the high-field limit 
is a fully spin-polarized set of single-electron orbitals 
with no nodal surfaces crossing the $z$ axis and with 
non-positive magnetic quantum numbers decreasing  
from $m=0$ to $m=-N+1$, where $N$ is the number of electrons. 
For the carbon atom, mentioned above, this Hartree-Fock configuration is 
$1s2p_{-1}3d_{-2}4f_{-3}5g_{-4}6h_{-5}$ with $S_z=-3$. 
For the sake of brevity we shall in the following refer 
to these ground state configurations 
in the high-field limit, i.e. the configuration generated by 
the tightly bound hydrogenic orbitals $1s, 2p_{-1}, 3d_{-2}, 4f_{-3},\ldots$, 
as $\left|0_N\right>$. 
The states $\left|0_N\right>$ possess the complete spin polarization $S_z=-N/2$. 
Decreasing the magnetic field strength, 
we can encounter a series of crossovers of the ground state configuration 
associated with transitions of one or several electrons 
from orbitals with the maximal values for $|m|$ to 
other orbitals with a different spatial geometry of the wave function 
but the same spin polarization. 
This means the first few crossovers can take place within the space of fully 
spin polarized configurations. 
We shall refer to these configurations by mentioning, i.e. noting, only the difference 
with respect to the state $\left|0_N\right>$. 
This notation can, of course, also be extended to non-fully spin 
polarized configurations. 
For instance the state $1s^22p_{-1}3d_{-2}4f_{-3}5g_{-4}$ with $S_z=-2$ of 
the carbon atom will be briefly refered to as $\left|1s^2\right>$, 
since the default is the occupation of the hydrogenic series 
$1s, 2p_{-1}, 3d_{-2},\ldots$ and 
only deviations from it are recorded by our notation.

In the following considerations we shall often refer to 
subsets of electronic states which possess different spin polarizations. 
As indicated above we will denote the set of electronic states 
with $S_z=-N/2$ as the FSP subset.  
Along with the global ground state it is expedient to consider 
also what we call local ground states which are the energetically 
lowest states with some definite degree of 
the spin polarization. 
For the purpose of the present work we need to know the local ground state 
of the subset of electronic states with $S_z=-N/2+1$ (which 
is the only partially spin polarized subset considered in this paper 
and which is refered to as subset PSP) 
in the high-field regime. 
This knowledge is necessary for the evaluation of the point of the crossover 
between the FSP and PSP ground states, i.e. 
for the determination of the critical field strengths at which 
the global ground state changes its spin polarization from $S_z=-N/2$ to $S_z=-N/2+1$. 
For sufficiently high fields the $\left|1s^2\right>$ state is the local ground state 
of the PSP subset of electronic states.

\section{Ground state electronic configurations in the 
high-field regime}

Let us start with the high field limit 
and the state $\left|0_N\right>$ and subsequently consider possible 
ground state crossovers which occur {\em with decreasing magnetic field strength}.
In the high-field regime we have per definition only crossovers due to changes 
of the spatial orbitals and no spin-flip crossovers. 
According to the goals of the present work we investigate the possible 
global ground state configurations belonging to the  
subset FSP and determine the transition 
points to the subset PSP. 
Since the detailed study of the latter subset of states 
for arbitrary field strengths goes beyond the scope of the present work 
we consider 
first only the $\left|1s^2\right>$ state of this subset 
which is the local ground state 
of the subset PSP for sufficiently strong fields.  
Then we investigate the FSP ground states 
with decreasing field strength until we reach  
the point of crossover with the energy of the configuration $\left|1s^2\right>$. 
Subsequently we need to consider other electronic configurations 
of the PSP set   
in order to determine the complete picture of the energy levels 
as a function of the field strength 
near the spin-flip crossover and, possibly, to correct the position of this point 
(the latter is necessary if the state $\left|1s^2\right>$ 
is not the lowest one of the subset PSP at the spin-flip point).

Let us consider the ground state transitions 
within the subset FSP with decreasing field strength. 
The first of these transitions occurs when the binding energy 
associated with the outermost orbital ($m_N=-N+1$) becomes less 
than the binding energy of one of the orbitals with $n_z>0$. 
Due to the circumstance, that all the orbitals with $n_z>0$ are 
not occupied in the high-field ground state configuration, 
it is reasonable to expect the transition of the outermost electron 
to one of the orbitals with $m=0$ and 
either $n_z=1$ (i.e. $2p_0$ orbital) or $n_z=2$ (i.e. $2s$ orbital). 
The decision between these two possibilities cannot be taken on the basis 
of qualitative arguments. 
For the hydrogen atom or hydrogen-like ions in a magnetic field 
the $2p_{0}$ orbital is more strongly bound than the
$2s$ orbital for any field strength. 
On the other hand, owing to the electronic screening 
of the nuclear charge in multi-electron atoms in field-free space
the $2s$ orbital tends to be more tightly bound than the $2p_0$ orbital.
Thus, we have two competing mechanisms and numerical calculations 
are required for the decision between the possible 
$\left|0_N\right> - \left|2p_0\right>$ and $\left|0_N\right> - \left|2s\right>$ 
crossovers to a new local FSP ground state. 
Our calculations for the $\left|2s\right>$ state 
presented below in table \ref{tab:2senergy} for neutral atoms 
and in table \ref{tab:2spenergy} for positive ions show 
that the state $\left|2s\right>$ becomes more tightly bound 
than the $\left|2p_0\right>$ state only for rather weak field strengths, 
where this state cannot pretend to be the ground state 
of the corresponding atom or ion 
due to the presence of more tightly bound non-fully spin polarized 
states. 
In result the first intermediate ground state of the subset 
FSP, i.e. the state beside the $\left|0_N\right>$ state which 
might be involved in the first crossover of the ground state 
with decreasing field strength, is the $\left|2p_0\right>$ state. 
Calculations for the subset PSP (see below) show indeed, 
that this state is the global ground state 
in a certain regime of field strengths 
for the neutral atoms with $Z\geq 6$, i.e. C, N, O, F and Ne, as well as 
their positive ions ${\rm C^+}$, ${\rm N^+}$, ${\rm O^+}$, ${\rm F^+}$, ${\rm Ne^+}$. 
For the atoms He, Li, Be and B ($Z\leq 5$) 
as well as for the ions ${\rm Li^+}$, ${\rm Be^+}$ and ${\rm B^+}$ 
the state $\left|1s^2\right>$ becomes more tightly bound than 
$\left|0_N\right>$ for fields stronger that those associated with 
the $\left|0_N\right> - \left|2p_0\right>$ crossover 
and the $\left|2p_0\right>$ never becomes the global ground state of these atoms and ions.
Thus, both neutral atoms and positive ions ${\rm A^+}$ with $Z\leq 5$ 
have only one fully spin polarized ground state configuration $\left|0_N\right>$, 
which represents the global ground state above some critical field strength.

The question about a possible second intermediate fully spin 
polarized ground state 
occurring with further decreasing field strength 
arises for neutral atoms and positive ions 
with $Z\geq 6$ which possess the intermediate fully spin polarized ground 
state $\left|2p_0\right>$. 
This state could be either a state, containing an additional 
orbital with $n_z=1$ 
which would result in the $\left|2p_03d_{-1}\right>$ configuration, 
or a state with an additional $s$-type orbital, i.e. $\left|2s2p_0\right>$.
The third possibility of the simultaneous transition 
of the electron with the magnetic quantum number $m_{N-1}=-N+2$ 
to the $3d_{-1}$ orbital and the electron in the $2p_0$ orbital to the $2s$ 
orbital, which gives the 
$\left|2s3d_{-1}\right>$ configuration, can 
be excluded from the list of possible 
ground state configurations without a numerical investigation. 
The reason herefore is 
that the $3d_{-1}$ orbital is for any field strength more weakly bound than 
the $2p_0$ orbital and thus the $\left|2s2p_0\right>$ configuration possess a 
lower energy than the $\left|2s3d_{-1}\right>$ configuration 
for arbitrary magnetic field strengths. 
When comparing the configurations $\left|2s2p_0\right>$ and $\left|2p_0 3d_{-1}\right>$ 
we can make use of what we have learned (see above) from the competing 
$\left|2p_0\right>$ and $\left|2s\right>$ configurations for higher field strengths:
The $2s$ orbital  
is energetically preferable at weak magnetic fields 
whereas the $3d_{-1}$ orbital yields energetically lower configurations 
in the strong field regime. 
Thus, we perform calculations for 
the $\left|2p_0 3d_{-1}\right>$ configuration 
for many field strengths 
and then perform at much fewer field strengths calculations 
to check the energy of the $\left|2s2p_0\right>$ configuration in order to obtain the correct 
lowest energy and state of the set FSP.

The behavior of the energy levels described in the previous paragraph 
is illustrated in Figure 1. 
In this figure the energy curves for four possible fully spin polarized electronic 
configurations and two energy curves for the PSP subset of the neon ($Z=10$) atom 
are presented. 
This figure shows, in particular, 
the energy curve of the high field ground state $\left|0_N\right>$ 
which intersects with the curve $E_{\left|2p_0\right>}(\gamma)$ at $\gamma=159.138$. 
The latter energy remains the lowest in the FSP subset until 
the intersection of this curve with $E_{\left|2p_0 3d_{-1}\right>}(\gamma)$ 
at $\gamma=40.537$. 
This intersection occurs at higher field strength than the intersection of 
the curves $E_{\left|2p_0\right>}(\gamma)$ and $E_{\left|1s^2\right>}(\gamma)$ 
which is at $\gamma=38.060$.
On the other hand, the control calculations for the state $\left|2s2p_0\right>$, 
not presented in Figure 1, show 
that its total energy for $\gamma=38.060$ is larger 
than the energy $E_{\left|2p_0 3d_{-1}\right>}$. 
According to the previous argumentation this means that the state 
$\left|2s2p_0\right>$ is not the global ground state of the Ne atom for any 
magnetic field strengths.
Furthermore the state $\left|2p_0 3d_{-1}\right>$ is a candidate 
for becoming the global ground state of the neon atom in some bounded regime 
of the field strength. 
However, we have not yet performed (see below) a detailed investigation 
of the lowest energy curves of the PSP subset which is essential 
to take a definite decision on the global ground state configurations. 
For neutral atoms with $6\leq Z \leq 9$ 
and positive ions ${\rm A^+}$ with $6\leq Z \leq 10$ 
the energies of the states $\left|2p_0 3d_{-1}\right>$ and $\left|2s2p_0\right>$ 
at the points of intersections of the curves 
$E_{\left|2p_0\right>}(\gamma)$ and $E_{\left|1s^2\right>}(\gamma)$ 
are higher than the energies of the states 
$\left|2p_0\right>$ and $\left|1s^2\right>$. 
This leads to the conjecture that no neutral atoms with $Z< 10$ 
and positive ions with $Z\leq 10$ 
can possess more than two different fully spin polarized ground state configurations 
in the complete range of field strengths.

The above concludes our considerations of the fully 
spin polarized ground state configurations. 
To prove or refute the above conjecture we have 
to address the question of 
the lower boundary  of the fully spin polarized domain, 
i.e. the lowest field strength, 
at which a fully spin-polarized state represents the ground 
state of the atom considered. 
It is evident that this boundary value of the field strength 
is given by the crossover from a fully spin polarized to 
a non-fully spin polarized ground state with decreasing field strength.

First of all we have to check if the state $\left|1s^2\right>$ has the 
lowest energy of all the states of subset PSP at the point 
of intersection of the curve $E_{\left|1s^2\right>}(\gamma)$ with 
the corresponding energy curve for the local ground state configuration 
of subset FSP. 
Following our considerations for the fully spin polarized case 
we can conclude that calculations have to be performed first of all 
for the states $\left|1s^2 2p_0\right>$ and $\left|1s^2 2s\right>$.

The numerical calculations show, that for atoms with $Z\leq 6$  and ions with $Z\leq 7$, 
the state $\left|1s^2\right>$ becomes 
the ground state while lowering the spin polarization 
from the maximal absolute value $S_z=-N/2$ to $S_z=-N/2+1$. 
For heavier atoms and ions we first remark that the state $\left|1s^2\right>$ 
is not the energetically lowest one in the PSP subset 
at magnetic fields at which 
its energy becomes equal to the energy of the lowest FSP state. 
For these atoms and ions the state $\left|1s^2 2p_0\right>$ lies lower 
than $\left|1s^2\right>$ at these field strengths. 
One can see this behavior for the neon atom in Figure 1.
The second possible PSP local ground state $\left|1s^2 2s\right>$ 
(not presented in Figure 1) proves to be less tightly bound at these fields. 
These facts allow in the following a definite clarification of 
the picture of the global ground state configuration 
in the high field regime. 
For atoms with $Z\geq 7$ and positive ions with $Z\geq 8$ the 
intersection points between the state $\left|1s^2 2p_0\right>$ and 
the energetically lowest state in the FSP subspace have to be calculated. 
In result, the spin-flip crossover occurs at higher fields than this would be 
in the case of $\left|1s^2\right>$ being the lowest state in the PSP subspace. 
In particular, the spin-flip crossover for the neon atom is found to be 
slightly higher than the point of the crossover $\left|2p_0\right>-\left|2p_0 3d_{-1}\right>$, 
and, therefore, this atom has in the framework of the Hartree-Fock approximation 
only two fully spin polarized configurations 
likewise other neutral atoms and positive ions with $6\leq Z \leq 10$. 
The above conjecture is therefore refuted and the FSP $\left|2p_0 3d_{-1}\right>$ 
represents never the global ground state configuration in the high field regime 
for all neutral atoms and positive ions with $Z\leq 10$. 
It should be noted that the situation with the neon atom can be regarded 
as a transient one due to closeness of 
the intersection $\left|2p_0\right>-\left|2p_0 3d_{-1}\right>$ 
to the intersection  $\left|2p_0\right>-\left|1s^2 2p_0\right>$. 
This means that we can expect the configuration $\left|2p_0 3d_{-1}\right>$ 
to be the global ground state for the sodium atom ($Z=11$). 
In addition an investigation of the neon atom carried out on 
a more precise level than the Hartree-Fock method could also introduce some corrections 
to the picture described above.

After obtaining the new spin flip points for atoms 
with $7\leq Z\leq 10$ and ions with $8 \leq Z\leq 10$ 
(which are transition points between the $\left|2p_0\right>$ and $\left|1s^2 2p_0\right>$
states) one has to check them with respect to the next 
(in the order of decreasing field strengths) possible PSP local ground state configurations. 
Analogously to the FSP subset these configurations are 
$\left|1s^2 2p_0 3d_{-1}\right>$ and $\left|1s^2 2s2p_0\right>$. 
The numerical calculations show, that their energies lie higher than the energy of 
the $\left|1s^2 2p_0\right>$ configuration at the spin flip points 
and they are therefore excluded from the list of the global ground states considered here.

The final picture of the crossovers 
of the global ground state configurations 
is presented in tables 
\ref{tab:atrans} (for the neutral atoms) and 
\ref{tab:ptrans} (for the positive ions $\rm{A^+}$). 
The corresponding values of the field strengths 
belonging to the point of crossover are underlined 
in these tables. 
The field strengths for other closelying crossovers 
which actually do not affect the scenario of the changes 
of the global ground state 
are also presented in these tables. 
In a graphical form these results are illustrated 
in Figures 2 (neutral atoms) and 3 (ions). 
Shown are the critical field strengths belonging to the crossovers 
of selected states of the atoms (ions) as functions of the nuclear charge. 
The filled symbols mark the crossovers of the energy levels 
which correspond to the actual transitions of the ground state configurations, 
whereas the analogous non-filled symbols correspond to magnetic field strengths 
of the crossovers not associated with changes in the ground state but excited states. 
One can see in these figures the dependencies of the field strengths for various types of 
crossovers on the charge of the nucleus. 
In particular, one can see many significant crossovers for $Z=10$ lying 
very close from each other on the $\gamma$ axis. 
This peculiarity in combination with the behavior 
of the curve $\gamma(Z)$ for the $\left|2p_0\right>-\left|2p_0 3d_{-1}\right>$ 
crossover allows one to expect the configuration $\left|2p_0 3d_{-1}\right>$ 
to become a ground state configuration for $Z>10$. 

Some summarizing remarks with respect to the global ground state configurations 
in the high field regime are in order. 
The atoms and positive ions with $Z\leq 5$ have one ground state configuration 
$\left|0_N\right>$. 
The atoms and ions with $6\leq Z\leq 10$ possess two high field configurations. 
The C atom ($Z=6$) plays an exceptional role in the sense that it is the only atom 
which shows the ground state crossover $\left|2p_0\right>-\left|1s^2\right>$ 
involving the $\left|1s^2\right>$ state as a global ground state.

\section{Numerical Results and Discussion}

The tables \ref{tab:grenergy}--\ref{tab:2spenergy} 
contain numerical values of the total energies 
of the neutral atoms and positive ions obtained in our Hartree-Fock calculations. 
Tables \ref{tab:grenergy}, \ref{tab:2p0energy}, \ref{tab:2s2energy} and \ref{tab:2senergy} 
contain the energies of the neutral atoms in the states 
$\left|0_N\right>$, $\left|2p_0\right>$, $\left|1s^2\right>$ and $\left|2s\right>$, 
respectively. 
The analogous results for the ions ${\rm A^+}$ are presented in tables 
\ref{tab:Apground}, \ref{tab:2p0Penergy}, \ref{tab:s2Penergy} and \ref{tab:2spenergy} 
(the results are for the states 
$\left|0_N\right>$, $\left|2p_0\right>$, $\left|1s^2\right>$ and $\left|2s\right>$). 
The energies associated with the points of crossover for the global ground state both in 
neutral atoms and in their singly positive ions are presented in table \ref{tab:trans}. 
These energy values provide us with the ionization energies at 
the transition points. 
Being combined with the data of the previous tables they provide 
the behavior of the ionization energies of the atoms and the
total energies of the atoms and positive ions in the complete high-field region.

In Figure 4 we present the ionization energies of neutral atoms divided by the 
ionization energy of the hydrogen atom as a function of the magnetic field strength. 
All the curves for multi-electron atoms at 
$\gamma <600$ lie lower than the curve for hydrogen at the corresponding field strengths. 
But for $\gamma>1500$ the ionization energies of all atoms 
exceed the ionization energy for the hydrogen atom. 
Moreover, with growing nuclear charge  we observe a stronger 
increase of the ionization energy for stronger fields accompanied by 
a shift of the starting point for the growth to the regime 
of stronger magnetic fields. 
This strengthening of the binding of the multi-electron atoms at 
strong magnetic fields may be considered as a hint for increasingly favorable 
conditions for the formation of the corresponding negative ions.

Figure 5 presents the ionization energies for the $\left|0_N\right>$ states 
for various field strengths 
depending on the nuclear charge $Z$, i.e. for all atoms H, He,\ldots, Ne. 
All the field strengths presented in this figure are 
above the first crossover to another global ground state configuration. 
Thus, the ionization energies in this figure represent the differences between 
the energies of the high-field ground states of the neutral atoms and 
the corresponding singly charged positive ions. 
The curve for $\gamma=2000$ can be considered as the prototype 
example for the general properties of the dependencies 
$E_{\rm Ion}(Z)$. 
For small values of $Z$ this curve shows increasing values for $E_{\rm Ion}$ 
with increasing $Z$, then it has a maximum at $Z=5$ and for $Z>5$ 
it decreases with increasing $Z$. 
Analogous curves for lower field strengths have their maxima 
at lower values of $Z$. 
At $\gamma=1000$ the ionization energy shows its maximal value at $Z=2$, 
whereas the ionization energies for $\gamma=500$ and $\gamma=200$ 
decrease monotonically with increasing $Z$. 
On the other hand, for $\gamma=5000$ and $\gamma=10000$ we obtain a  
monotonically increasing behavior of the ionization energy 
for the whole range $1\leq Z\leq 10$ 
of nuclear charges investigated in the present work. 
The behavior described above results from a competition of two different 
physical mechanisms which impact the binding energy of the outermost 
electron in the high-field ground state Hartree-Fock configuration. 
The first mechanism is the lowering of the binding energy 
of the outermost electron with increasing absolute value of its 
magnetic quantum number $|m|$ provided that this electron feels 
a constant nuclear charge. 
The latter assumption is a rough approximation to the case of relatively 
weak fields when the inner $Z-1$ electrons screen more or less effectively 
the Coulomb field of the nucleus. 
The second and opposite tendency is associated with the decrease of 
the efficiency of this screening in extremely strong magnetic fields due to 
the fact that the geometry of the wave functions tends to be one-dimensional 
in these fields. 
In result the effect of increasing effective nuclear charge exceeds the effect 
of the growth of $|m|$ with increasing $Z$ 
for the high-field ground state configurations. 
Continuing this qualitative consideration we point out that at each 
fixed $\gamma$ the influence of the magnetic field on the inner electrons 
become less and less significant as $Z$ increases which is due 
to the dominance of the Coulomb attraction potential of the nucleus 
over the magnetic field interaction.
This has to result in a significant screening of the nuclear charge by 
these electrons. 
In result  
the functions $E_{\rm Ion}(Z)$ for strong fields 
defined on the whole interval $1\leq Z < +\infty$ 
have maxima at some values for $Z$ and 
decrease for sufficiently large values of $Z$. 

Next we provide a comparison of the present results 
with adiabatic HF calculations which were carried out for multi-electron atoms 
in refs. \cite{Neuhauser,Godefroid}. 
We compare our results on the Hartree-Fock electronic structure 
of atoms in strong magnetic fields with results obtained by 
Neuhauser et al \cite{Neuhauser} via a one-dimensional {\it adiabatic} 
Hartree-Fock approximation. 
The calculations in this work were carried out for the four field 
strengths $\gamma=42.544$,  $\gamma=212.72$, $\gamma=425.44$ and $\gamma=2127.2$. 
For $Z\leq 9$ and all these field strengths and for $Z=10$ at the three larger values  
of these fields the Hartree-Fock wave functions of the ground states are 
reported to be fully spin polarized with no nodes crossing the $z$ axis. 
This conclusion differs from our result for $\gamma=42.544$. 
According to our calculations at $\gamma=42.544$ 
the wave functions without nodes crossing the $z$ axis
represent the ground states of atoms with $Z\leq 7$ 
(i.e. H, He, Li, Be, B, C and N) whereas for the atoms with 
$8\leq Z \leq 10$ (i.e. O, F and Ne) the wave functions of the ground 
states are fully spin polarized 
with one nodal surface crossing the $z$ axis. 
A numerical comparison of our results with those of refs. \cite{Neuhauser,Godefroid} 
is shown in table \ref{tab:Neucomp}. 
All our values lie lower than the values of these adiabatic calculations. 
Since our total energies are upper bonds to the exact values 
we consider our HF results as being closer to the exact values compared to the results of 
the adiabatic HF calculations. 
Therefore, on the basis of our calculations combined with the results of \cite{Neuhauser,Godefroid} 
we can obtain an idea of the degree of the applicability of the adiabatic approximation 
for multi-electron atoms for different field strengths and nuclear charges. 
It is well known, that the precision of the adiabatic approximation decreases with 
decreasing field strength. 
The increase of the relative errors with decreasing field strength is clearly 
visible in the table. 
On the other hand, the relative errors of the adiabatic approximation 
possess the tendency to increase with growing $Z$, which is manifested 
by the scaling transformation $E(Z,\gamma)=Z^2 E(1,\gamma/Z^2)$ 
(e.g. \cite{Rud94,Ivanov94}) well known for hydrogen-like ions. 
The behavior of the inner electrons is to some extent similar to the behavior 
of the electrons in the corresponding hydrogen-like ions. 
Therefore their behavior is to lowest order similar to the behavior of the electron in 
the hydrogen atom at magnetic field strength $\gamma/Z^2$ i.e. this 
behavior can be less accurately described by 
the adiabatic approximation at large $Z$ values. 
The absolute values of the errors in the total energy associated with the adiabatic approximation 
are in many cases larger than the corresponding values of the ionization energies.

To conclude this section we discuss briefly 
three issues, which could affect the precision of the results presented above.
These issues are electron correlations, 
effects due to the finite nuclear mass and relativistic corrections. 
For all these effects we have to distinguish between their influence on the total 
energy and on other quantities like the ionization energy and 
the field strength for the crossover of the energy levels. 
In most cases their influence on the latter values is much smaller 
due to the fact that they involve differences of total energies 
for quantum states possessing a similar atomic core. 
Let us start by addressing the problem of the electronic correlations which 
is the critical problem 
for the precision of the Hartree-Fock calculations. 
The final evaluation of the correlation effects is possible only on 
the basis of exact calculations going beyond the Hartree-Fock approximation. 
Therefore we can give here only qualitative arguments based on the geometry 
of the wave function and on existing 
calculations for less complicated systems.  
The dependence of the ratio of the correlation energy and the total 
binding energy for the two ground state configurations of the helium atom 
has been investigated in ref. \cite{Schm98b}. 
This ratio for the $1s^2$ state decreases with growing $\gamma$ from 
$1.4\%$ at $\gamma=0$ to about $0.6\%$ at $\gamma=100$. 
The same ratio for the $1s2p_{-1}$ state (high field ground state configuration) 
increases with growing $\gamma$. 
It remains however for all the field strengths considered 
essentially smaller the values for the $1s^2$ state. 
This result for the helium atom in strong magnetic fields allows us 
to speculate that for the field strengths considered here 
the correlation energy for atoms and positive ions heavier 
than helium atom does not exceed their corresponding values 
without fields. 
Due to the similar geometry of the inner shells in the participating 
electronic configurations we do not expect a major influence of 
the correlation effects both on the field strengths 
of the crossovers of the ground state 
configurations within the subsets FSP or PSP and on 
the ionization energies if the states of a neutral atom 
and the positive ion belong to the same subset. 
On the other hand, the properties associated with configurations 
from different subsets (for instance values of the spin-flip crossover field strengths) 
can be affected more strongly by correlation effects. 

Our second issue is the influence of the finite nuclear mass on the results 
presented above. 
A discussion of this problem is provided in ref. \cite{Bec99} and references therein. 
Importantly there exists a well-defined procedure which tells us how to relate 
the energies for infinite nuclear mass to those with a finite nuclear mass. 
The corresponding equations are exact for hydrogen-like systems and 
provide the lowest order mass corrections $O\left(\frac m M \right)$ 
($m$ and $M$ are the electron and total mass, respectively) for general atoms/ions. 
Essentially they consist of a redefinition of the energy scale (atomic units $\longrightarrow$ 
reduced atomic units, due to the introduction of the reduced mass) 
and an additional energy shift $-(1/M_0)\gamma(M+S_z)$ 
where $M_0$ is the nuclear mass. 
The first effect can simply be 'included' in our results by taking 
the energies in reduced a.u. instead of a.u. 
The mentioned shift can become relevant for high fields. 
However, it can easily be included in the total energies 
presented here. 
We emphasize that it plays a minor role 
in the region of the crossovers of the ground state configurations 
and decreases significantly with increasing mass of the atom (nucleus).

Relativistic calculations for the hydrogen atom and hydrogen-like ions were performed 
by Lindgren and Virtamo \cite{LindgrenVirt} and Chen and Goldman \cite{ChenGoldRelat}. 
Our considerations are based on the work by Chen and Goldman \cite{ChenGoldRelat} 
which contains results for the $1s$ and $2p_{-1}$ states for a broad range of 
magnetic field strengths. 
Interpolating their results for the $1s$ state and using well known scaling transformations 
we can conclude that in the least favorable case of $Z=10$ 
relativistic corrections 
$\delta E =~ (E^{\rm relativistic}-~E^{\rm non-relativistic})/~|E^{\rm non-relativistic}|$ 
have to be of the order $4\cdot 10^{-4}$ for $\gamma=200$ 
and $2\cdot 10^{-4}$ for $\gamma=10^4$. 
The relativistic corrections for the $2p_{-1}$ state at relatively strong 
fields appear to be of the same order of magnitude or smaller than for the $1s$ state. 
Thus, making a reasonable assumption 
that relativistic corrections for both inner and outer 
electrons are similar to those in the hydrogen-like ions with a properly scaled 
nuclear charge we can evaluate $|\delta E| \leq 4\cdot 10^{-4}$ for $Z=10$ 
and lesser for lower $Z$ values. 
The same relative 
correction can be expected also for the ionization energies and energy values 
used for the determination of the crossovers of the electronic configurations.

\section{Summary}

In the present work we have applied our 2D Hartree-Fock method to the magnetized neutral atoms 
H, He, Li, Be, B, C, N, O, F and Ne 
in the high field regime which is characterized by 
fully spin-polarized electronic shells. 
Additionally we have studied the crossover from fully spin polarized to partially 
spin polarized global ground state configurations. 
The highest field strength investigated was $\gamma=10000$. 
Our single-determinant Hartree-Fock approach supplies us with 
exact upper bounds for the total energy. 
A comparison with adiabatic calculations in the literature shows 
the decrease of the precision of the latter with growing $Z$. 

The investigation of the geometry of the spatial part 
of the electronic wave function demonstrates that 
in the high-field limit this wave function is a composition of the lowest Landau 
orbitals with absolute values of the magnetic quantum number growing from $|m|=0$ up to 
$|m|=N-1$, where $N$ is the number of the electrons: 
i.e. we have the series $1s$, $2p_{-1}$, $3d_{-2}$,\ldots 
For atoms with $2 \leq Z \leq 5 $ these states of type $1s2p_{-1}3d_{-2}\ldots$ represent 
the complete set of the fully spin-polarized ground state configurations. 
Heavier atoms $6 \leq Z \leq 10$ have one intermediate ground state configuration 
associated with the low-field end of the fully spin polarized region. 
This state contains one $2p_0$ type orbital 
(i.e. the orbital with a negative $z$ parity and $|m|=0$) 
instead of the orbital with the positive $z$ parity and the maximal value of $|m|$. 
Extrapolating our data as a function of the nuclear charge $Z$ 
we expect that a third fully spin polarized ground state configuration 
occurs first for $Z=11$, i.e. the sodium atom. 
The third configuration is suggested to be the $\left|2p_0 3d_{-1}\right>$ state. 
The critical field strength which provides the crossover 
from the partially spin polarized to the fully spin polarized 
regime depends sensitively on the changes of the geometry 
of the wave functions. 
Indeed a number of different configurations have been selected 
as candidates for ground states in the crossover regime and 
only concrete calculations could provide us with a final decision on 
the energetically lowest state of the non-fully spin polarized 
electronic states. 
Generally speaking all the spin-flip crossovers mentioned above involve 
a pairing of the $1s$ electrons, i.e. the pair of orbitals $1s^2$. 
The carbon atom ($Z=6$) plays an exceptional role since it is the only 
neutral atom which possesses two fully spin polarized configurations and 
the $\left|1s^2\right>$ as a global non-fully spin polarized ground state configuration. 
The spin-flip crossover of the carbon atom preserves the total magnetic quantum number. 
All other atoms N, O, F and Ne ($7\leq Z\leq 10$) possess instead the 
$\left|1s^2 2p_0\right>$ configuration as a non-fully spin polarized ground state 
for strong fields. 
We have determined the positions, i.e. field strengths, of the crossovers of 
the ground states. 
Beyond this total energies have been provided for many field strengths for 
several low-lying excited states.

An analogous investigation has been carried out for singly charged positive ions $2 \leq Z \leq 10$. 
The structure of the fully spin polarized ground state configurations for these ions is the following: 
The ions with $3 \leq Z \leq 5$ have one fully spin polarized ground state configuration analogous to 
the high-field limit of the neutral atoms. 
For $6 \leq Z \leq 10 $, analogously to the neutral atoms, 
there exist two fully spin polarized ground state configurations. 
Depending on the values of the nuclear charge number $Z$ 
the spin-flip transitions associated with the lowering 
of the spin polarization with decreasing field strength 
lead also to wave functions of different spatial symmetries. 
These data being combined with the data for neutral atoms 
allow us to obtain the ionization energies of the atoms. 
The dependencies of the ionization energies on the nuclear charge at fixed field strength 
generally exhibit maxima at certain values of $Z$. 
The positions of these maxima shift to larger values of $Z$ with increasing field strength. 
We provide some qualitative arguments explaining this behavior of $E_{\rm Ion}(Z)$. 
Finally we have given some remarks on the interactions 
going beyond the present level of investigation, i.e. correlations 
and finite nuclear mass effects as well as relativistic corrections. 

\vspace*{0.5cm}

\begin{center}
{\bf{Acknowledgments}}
\end{center}
Financial support by the
Deutsche Forschungsgemeinschaft is gratefully acknowledged. 
\newpage


{}

\vspace*{2.0cm}

{\bf Figure Captions}

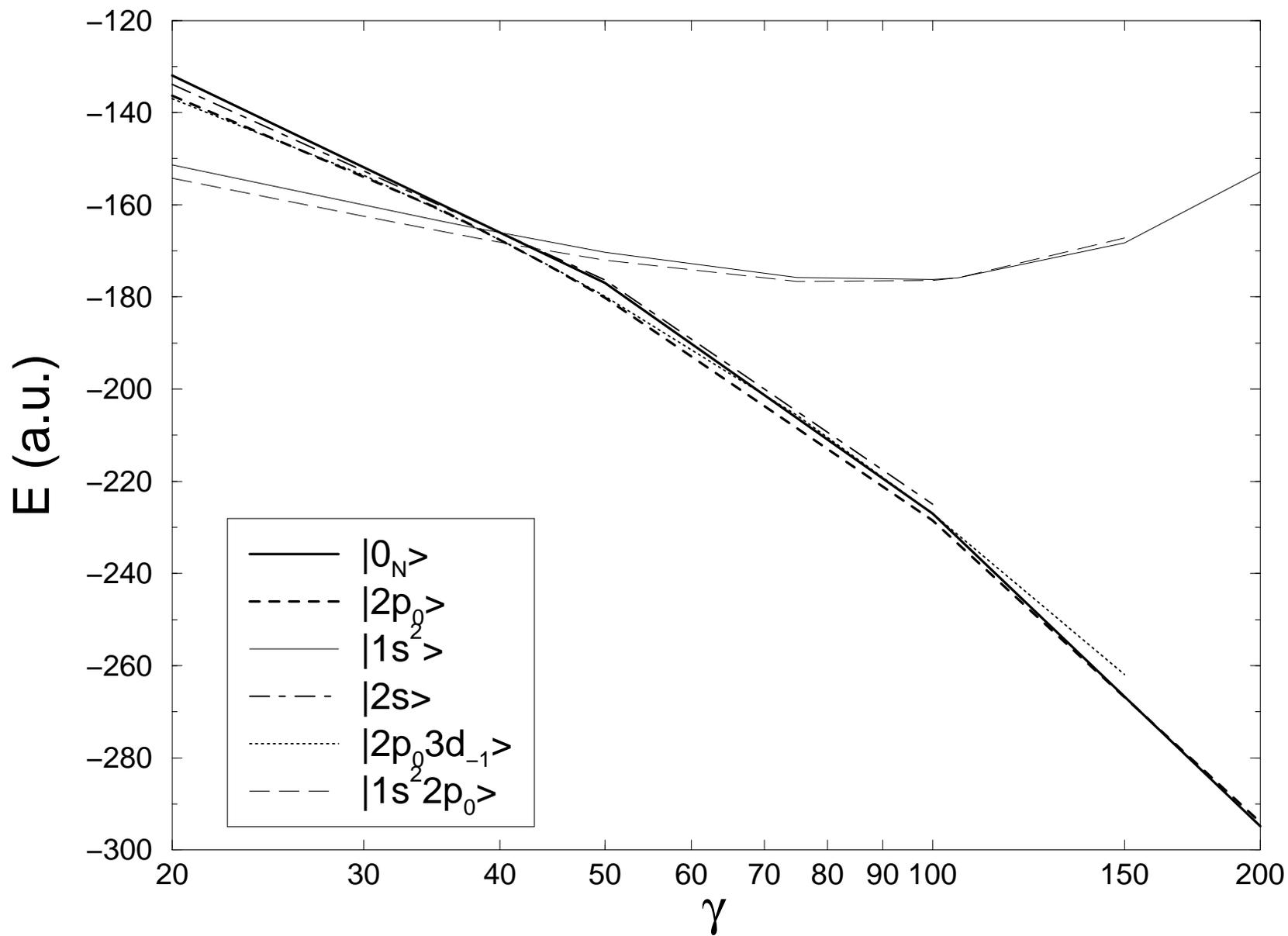
{\bf Figure 1.} The total energies (in atomic units) of the relevant states of 
the neon atom under consideration for the determination of the ground state 
electronic configurations for the high field regime. 

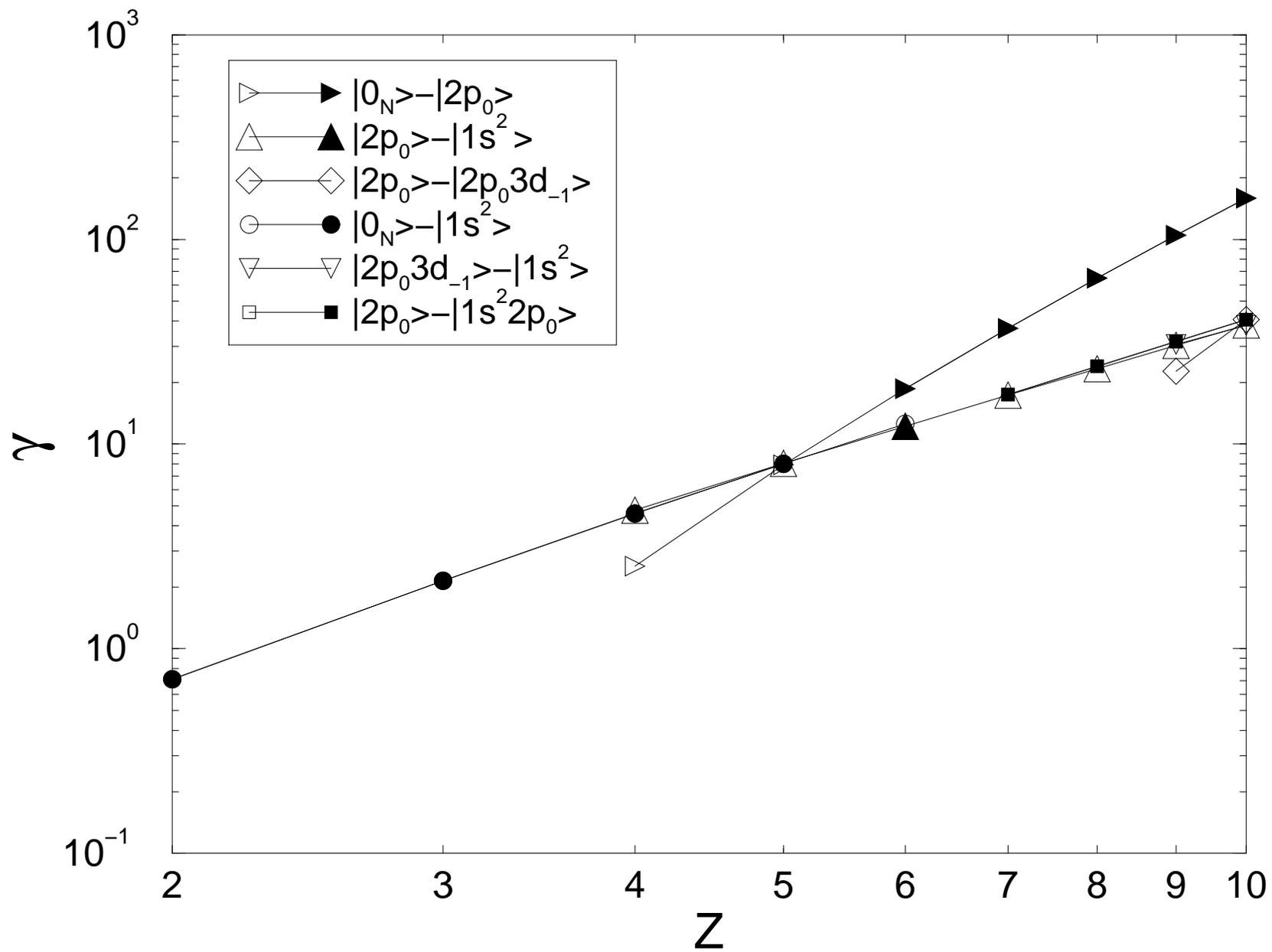
{\bf Figure 2.} The magnetic field strengths (a.u.) corresponding 
to crossovers of energy levels 
in neutral atoms as functions of the nuclear charge. 
The filled symbols mark crossovers between global ground state configurations. 

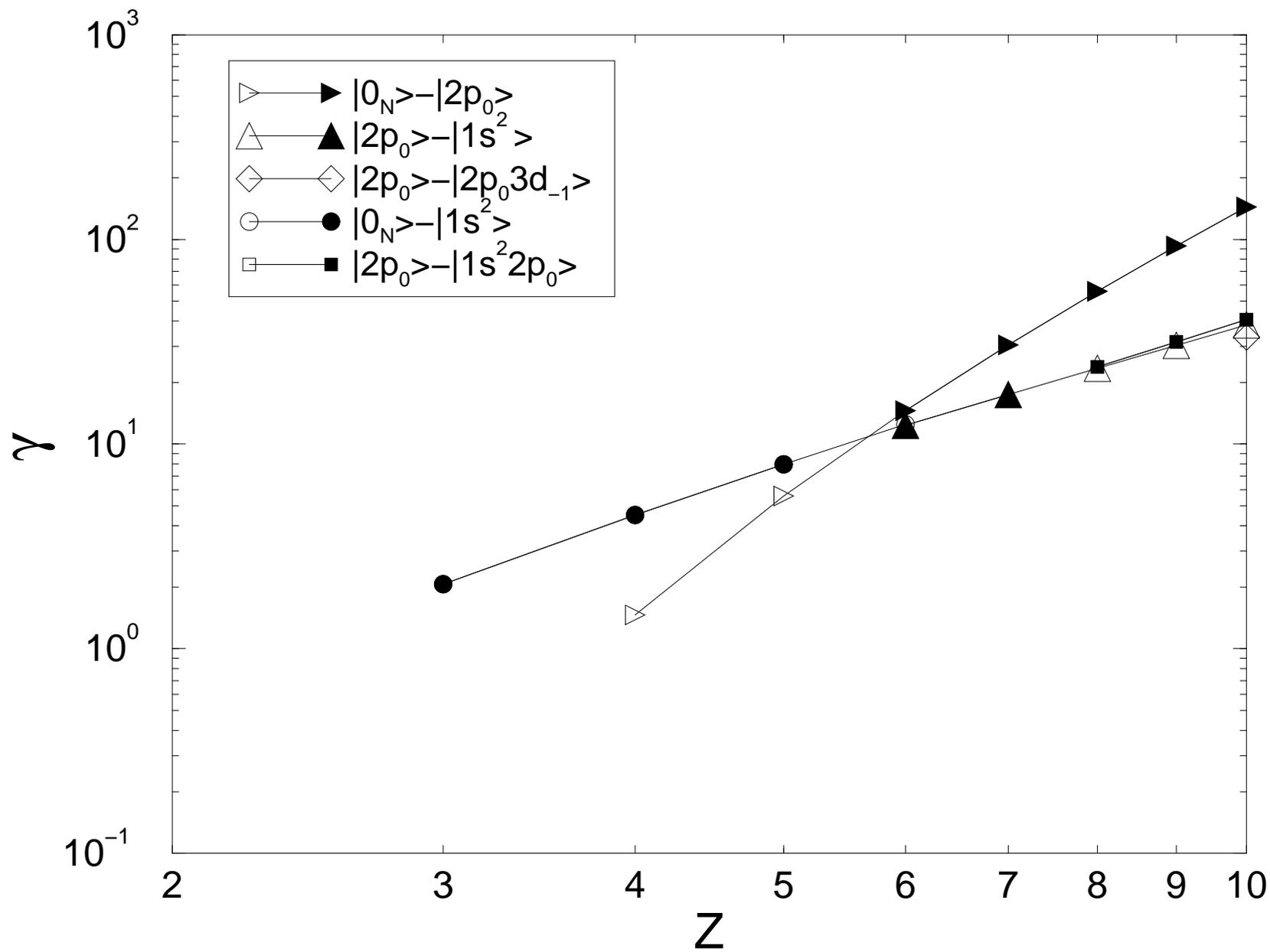
{\bf Figure 3.} Same as Figure 2 but for the singly positive ions. 

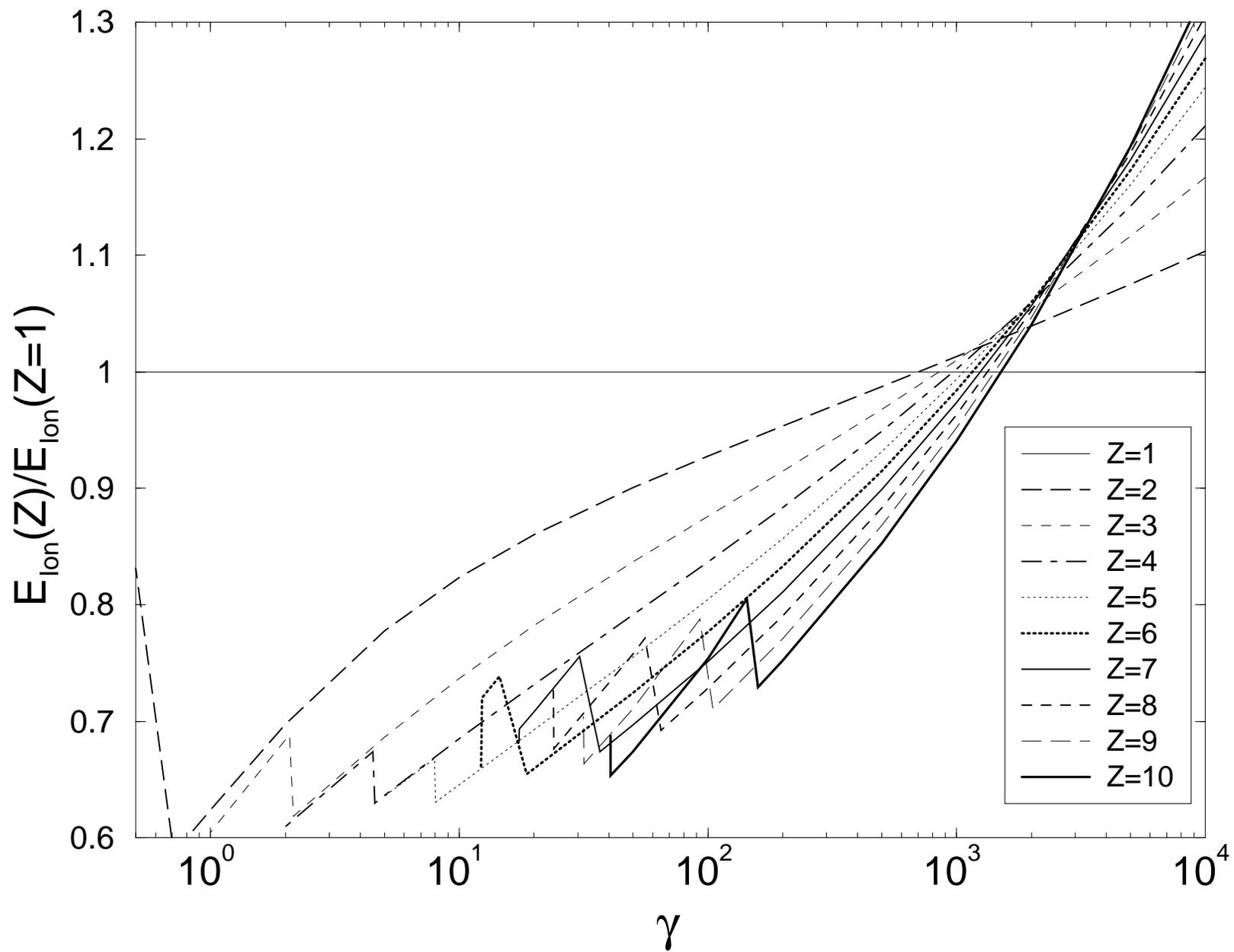
{\bf Figure 4.} Ionization energies of neutral atoms divided by the 
ionization energy of the hydrogen atom as a function of the magnetic field strength (a.u.). 

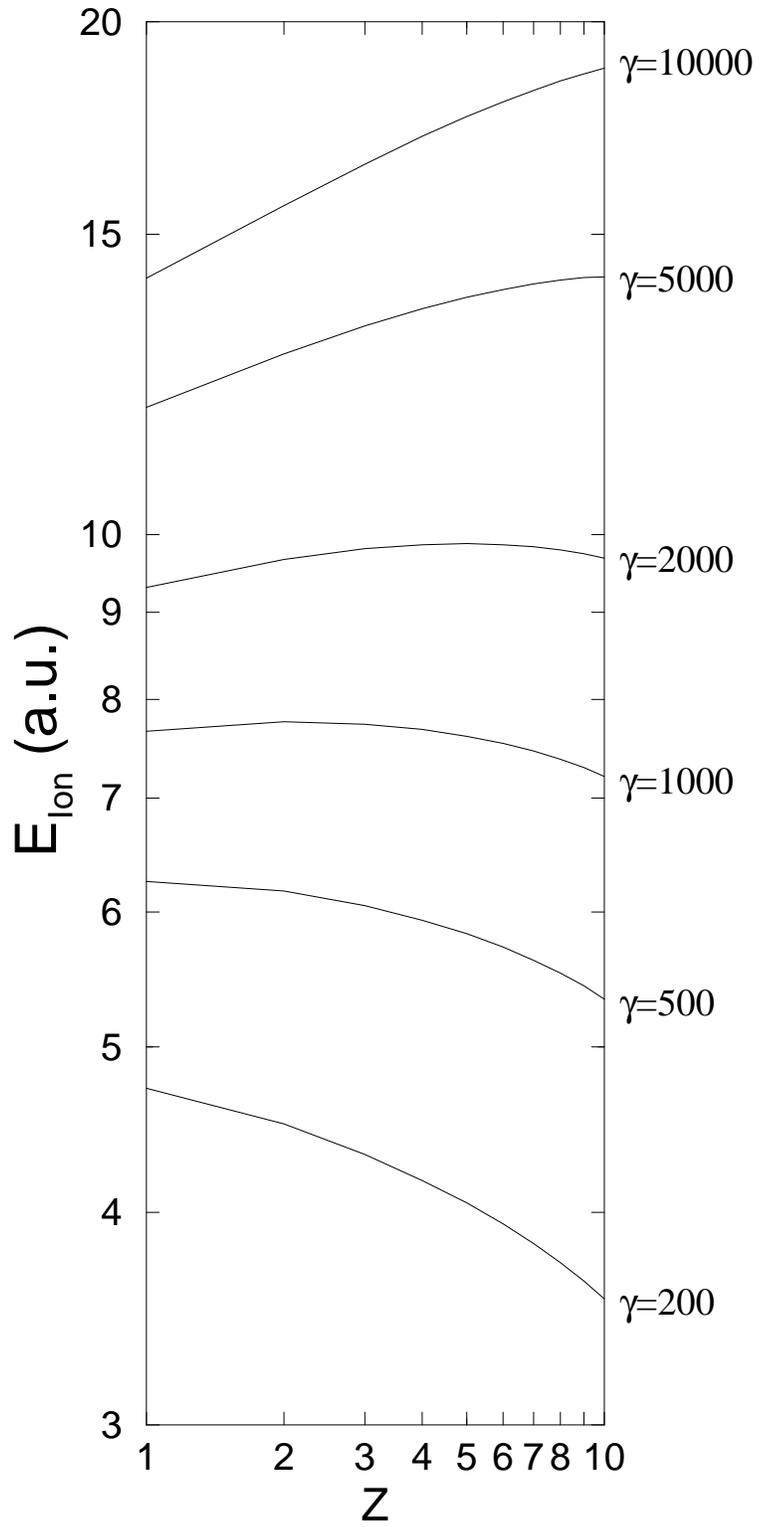
{\bf Figure 5.} Ionization energies of the states $\left|0_N\right>$ 
of the neutral atoms ($1\leq Z\leq 10$) 
for different magnetic field strengths.

\newpage

\begin{table}
\caption{Magnetic field strengths $\gamma$ (a.u.) for energy level crossovers in neutral atoms. 
Ground state crossovers are underlined.}
\begin{tabular}{@{}lllllllll}
Z  &$\left|0_N\right>-\left|1s^2\right>$&$\left|0_N\right>-\left|2p_0\right>$&$\left|2p_0\right>-\left|1s^2\right>$&$\left|2p_0\right>-\left|1s^22p_0\right>$&$\left|2p_0\right>-\left|2p_03d_{-1}\right>$&$\left|2p_03d_{-1}\right>-\left|1s^2\right>$\\ \noalign{\hrule}
2  &\underline{  0.711}    &$           	  $&$         $&                  &$                $&$                $\\
3  &\underline{  2.153}    &$           	  $&$         $&                  &$                $&$                $\\
4  &\underline{  4.567}    &\ \     2.529	   &\ 4.765451 &                  &$                $&$                $\\
5  &\underline{  8.0251}   &\ \     7.923	   &\ 8.0325   &                  &$                $&$                $\\
6  &$  12.577             $&\underline{ 18.664}	   &\underline{12.216}   &        &$                $&$                $\\
7  &$                     $&\underline{ 36.849}	   &{17.318}   &\underline{17.398}&$                $&$                $\\
8  &$                     $&\underline{ 64.720}	   &{23.3408}  &\underline{23.985}&$                $&$                $\\
9  &$                     $&\underline{104.650}	   &{30.285 }  &\underline{31.735}&$22.744          $&$30.6125         $\\
10 &$                     $&\underline{159.138}	   &38.151     &\underline{40.672}&40.537            &38.060\\
\end{tabular}
\label{tab:atrans}
\end{table}

\begin{table}
\caption{Magnetic field strengths $\gamma$ (a.u.) for energy level crossovers in positive ions $\rm A^+$.  
Ground state crossovers are underlined.}
\begin{tabular}{@{}lllllllll}
Z  &$\left|0_N\right>-\left|1s^2\right>$&$\left|0_N\right>-\left|2p_0\right>$&$\left|2p_0\right>-\left|1s^2\right>$&$\left|2p_0\right>-\left|1s^22p_0\right>$&$\left|2p_0\right>-\left|2p_03d_{-1}\right>$\\ \noalign{\hrule}
3  &\ \underline{2.0718}   &$           	  $&$                $&$                $&$	         $\\
4  &\ \underline{4.501 }   &$\ \ 1.464 	          $&$                $&$                $&$	         $\\
5  &\ \underline{7.957 }   &$\ \ 5.575 	          $&$                $&$                $&$	         $\\
6  &$12.506		  $&\ \underline{14.536}   &\underline{12.351}&                  &$	         $\\
7  &$			  $&\ \underline{30.509}   &\underline{17.429}&                  &$	         $\\
8  &$			  $&\ \underline{55.747}   &23.434            &\underline{23.849}&$	         $\\
9  &$			  $&\ \underline{92.624}   &30.364            &\underline{31.612}&$	         $\\
10 &$			  $&\underline{143.604}	   &38.220            &\underline{40.559}&$33.353        $\\
\end{tabular}
\label{tab:ptrans}
\end{table}

\newpage
\begin{table}
\caption{Total energies (a.u.) of the high-field ground states $\left|0_N\right>$ 
of neutral atoms in strong magnetic fields.}
\begin{tabular}{@{}rlllllllll}
$Z   $&$\gamma=0.5$&$\gamma=1   $&$\gamma=2   $&$\gamma=5   $&$\gamma=10  $&$\gamma=20  $&$\gamma=50  $&\\ \noalign{\hrule}  
1     &$-0.69721056$&$-0.83116892$&$-1.0222139 $&$-1.38039889$&$-1.74779718$&$-2.21539853$&$-3.01786074$&\\
2     &$-2.615551 $&$-2.959690	$&$-3.502051  $&$-4.617251  $&$-5.829513  $&$-7.427704  $&$-10.264493 $&\\
3     &$-5.97052  $&$-6.57080	$&$-7.520029  $&$-9.576936  $&$-11.939018 $&$-15.1626119$&$-21.05055  $&\\
4     &$-10.80902 $&$-11.72880	$&$-13.16961  $&$-16.30690  $&$-20.01753  $&$-25.232499 $&$-35.00768  $&\\
5     &$-17.1771  $&$-18.45812	$&$-20.46843  $&$-24.83956  $&$-30.06363  $&$-37.55469  $&$-51.91499  $&\\ 
6     &$-25.1007  $&$-26.7843	$&$-29.4282   $&$-35.18153  $&$-42.07989  $&$-52.08903  $&$-71.6285   $&\\
7     &$-34.5971  $&$-36.7230	$&$-40.0600   $&$-47.3314   $&$-56.06309  $&$-68.81304  $&$-94.0501   $&\\
8     &$-45.6798  $&$-48.2846   $&$-52.3718   $&$-61.2866   $&$-72.005397 $&$-87.7104   $&$-119.112   $&\\ 
9     &$-58.3588  $&$-61.4777	$&$-66.3692   $&$-77.0449   $&$-89.89720  $&$-108.7661  $&$-146.7620  $&\\ 
10    &$	  $&$		$&$	      $&$-94.60624  $&$-109.7289  $&$-131.9650  $&$-176.964   $&\\
\noalign{\hrule}  
\noalign{\hrule}  
\noalign{\hrule}  
\noalign{\hrule}  
$Z   $&$\gamma=100$&$\gamma=200$&$\gamma=500 $&$\gamma=1000$&$\gamma=2000$&$\gamma=5000$&$\gamma=10000$&\\ \noalign{\hrule}  
1     &$-3.7898043$&$-4.7271451$&$-6.257088  $&$-7.6624234 $&$-9.3047652 $&$-11.873419 $&$-14.14097 $&\\
2     &$-13.07665 $&$-16.57908 $&$-22.46665  $&$-28.03209  $&$-34.6989   $&$-45.4246   $&$-55.1514  $&\\
3     &$-27.01927 $&$-34.58499 $&$-47.55830  $&$-60.05892  $&$-75.282411 $&$-100.2482  $&$-123.313  $&\\
4     &$-45.10519 $&$-58.08264 $&$-80.67357  $&$-102.75480 $&$-129.9790  $&$-175.2704  $&$-217.695  $&\\
5     &$-66.99699 $&$-86.60738 $&$-121.16488 $&$-155.3296  $&$-197.8655  $&$-269.440   $&$-337.230  $&\\ 
6     &$-92.4552  $&$-119.8127 $&$-168.5248  $&$-217.1413  $&$-278.1612  $&$-381.8097  $&$-480.875  $&\\
7     &$-121.3027 $&$-157.4300 $&$-222.3434  $&$-287.65764 $&$-370.2004  $&$-511.536   $&$-647.685  $&\\
8     &$-153.405  $&$-199.2455 $&$-282.28330 $&$-366.430   $&$-473.413   $&$-657.871   $&$-836.767  $&\\
9     &$-188.657  $&$-245.085  $&$-348.0593  $&$-453.0748  $&$-587.294   $&$-820.140   $&$-1047.3242$&\\ 
10    &$-226.976  $&$-294.807  $&$-419.430   $&$-547.259   $&$-711.4106  $&$-997.7478  $&$-1278.622 $&\\
\end{tabular}
\label{tab:grenergy}
\end{table}

\begin{table}
\caption{Total energies (a.u.) of neutral atoms in strong magnetic fields 
in the fully spin polarized states $\left|2p_0\right>$.}
\begin{tabular}{@{}rllllllllllll}
$Z   $&$\gamma=0.5 $&$\gamma=1   $&$\gamma=2   $&$\gamma=5  $&$\gamma=10  $&$\gamma=20  $&$\gamma=50  $&$\gamma=100 $&$\gamma=200 $\\ \noalign{\hrule}
1     &$-0.224760$&$-0.260007$&$-0.297711$&$-0.347618$&$-0.382650$&$-0.413378$&$-0.445685$&$-0.463618$&$-0.476532$\\
2     &$-2.477333  $&$-2.730171 $&$-3.130766  $&$-3.953993 $&$-4.842630   $&$-6.00481   $&$-8.05248   $&$-10.072    $&$-12.588	$\\
3     &$-5.969573  $&$-6.492478	 $&$-7.324937  $&$-9.125540 $&$-11.17884  $&$-13.96583  $&$-19.0436   $&$-24.1951   $&$-30.734	$\\
4     &$-11.06254  $&$-11.89891	 $&$-13.22133  $&$-16.10812 $&$-19.51207  $&$-24.27725  $&$-33.2000   $&$-42.4440   $&$-54.368	$\\
5     &$	   $&$-19.05098	 $&$-20.92634  $&$-25.03513 $&$-29.94166  $&$-36.95414  $&$-50.377973 $&$-64.5298   $&$-83.031	$\\
6     &$           $&$-28.0195	 $&$-30.4938   $&$-35.96012 $&$-42.52774  $&$-52.02820  $&$-70.51870  $&$-90.275    $&$-116.4070$\\
7     &$	   $&$-38.8370	 $&$-41.9590   $&$-48.9040  $&$-57.29256  $&$-69.5147   $&$-93.6004   $&$-119.5977  $&$-154.272 $\\
8     &$	   $&$-51.5182	 $&$-55.3413   $&$-63.877   $&$-74.2380   $&$-89.4093   $&$-119.592   $&$-152.453   $&$-196.522 $\\ 
9     &$	   $&$-66.0734	 $&$-70.6514   $&$-80.8826  $&$-93.3580   $&$-111.6968  $&$-148.4508  $&$-188.7802  $&$-243.1024$\\ 
10    &$	   $&$-82.5108	 $&$-87.8960   $&$-99.9271  $&$-114.64655 $&$-136.36054 $&$-180.1312  $&$-228.500   $&$-293.944 $\\
\end{tabular}
\label{tab:2p0energy}
\end{table}

\begin{table}
\caption{Total energies (a.u.) of neutral atoms in strong magnetic fields 
in the states $\left|1s^2\right>$.}
\begin{tabular}{@{}rllllllllllll}
$Z   $&$\gamma=0.5$&$\gamma=1 $&$\gamma=2 $&$\gamma=5 $&$\gamma=10$&$\gamma=20 $&$\gamma=50 $&$\gamma=100$&$\gamma=200$&\\ \noalign{\hrule}
2     &$-2.8144511$&$-2.688885$&$-2.289145$&$-0.532445$&$+3.110634$&$+11.319608$&$+38.14390 $&$+85.00416 $&$+181.10639$&\\
3     &$-7.58789  $&$-7.666532$&$-7.662455$&$-6.942304$&$-4.617769$&$+1.705656 $&$+24.97942 $&$+68.17347 $&$+159.57479$&\\
4     &$-14.82273 $&$-15.16179$&$-15.57496$&$-15.91027$&$-15.04644$&$-10.97100 $&$+7.83395  $&$+46.25962 $&$+131.4188 $&\\
5     &$-24.5395  $&$-25.20257$&$-26.11859$&$-27.59737$&$-28.27946$&$-26.68603 $&$-13.06555 $&$+19.65113 $&$+97.1970  $&\\ 
6     &$-36.7864  $&$-37.8130 $&$-39.3061 $&$-42.06081$&$-44.38721$&$-45.44649 $&$-37.57176 $&$-11.36933 $&$+57.3384  $&\\
7     &$-51.5899  $&$-53.0202 $&$-55.1513 $&$-59.3169 $&$-63.4083 $&$-67.27185 $&$-65.5935  $&$-46.5970  $&$+12.1743  $&\\
8     &$-68.967	  $&$-70.8400 $&$-73.6672 $&$-79.3704 $&$-85.3599 $&$-92.1817  $&$-97.0777  $&$-85.8840  $&$-38.0379  $&\\ 
9     &$-88.930	  $&$-91.2830 $&$-94.8623 $&$-102.2227$&$-110.2464$&$-120.1897 $&$-131.9955 $&$-129.1244 $&$-93.0956  $&\\ 
10    &$-111.491  $&$-114.3575$&$-118.7427$&$-127.8738$&$-138.0661$&$-151.3018 $&$-170.3322 $&$-176.2422 $&$-152.8395 $&\\
\end{tabular}
\label{tab:2s2energy}
\end{table}

\begin{table}
\caption{Total energies (a.u.) of neutral atoms in strong magnetic fields 
in the states $\left|2s\right>$.}
\begin{tabular}{@{}rllllllllllll}
$Z   $&$\gamma=0.5$&$\gamma=1 $&$\gamma=2 $&$\gamma=5 $&$\gamma=10$&$\gamma=20 $&$\gamma=50 $&$\gamma=100$\\ \noalign{\hrule}
2     &$-2.452834 $&$-2.649185$&$-2.998243$&$-3.76667 $&$-4.62593 $&$-5.7711   $&$-7.8134   $&$-9.8438	 $\\
3     &$-6.047868 $&$-6.480293$&$-7.188888$&$-8.88983 $&$-10.91060$&$-13.69420 $&$-18.8014  $&$-23.9861	 $\\
4     &$-11.23262 $&$-11.99646$&$-13.14233$&$-15.78294$&$-19.12479$&$-23.89990 $&$-32.9045  $&$-42.2253	 $\\
5     &$-18.1278  $&$-19.24491$&$-20.95537$&$-24.62942$&$-38.19566$&$-36.35453 $&$-49.9355  $&$-64.243	 $\\ 
6     &$	  $&$	      $&$	  $&$-35.57332$&$-41.68450$&$-51.0639  $&$-69.770585$&$-89.815	 $\\
7     &$	  $&$	      $&$	  $&$-48.6234 $&$-56.2802 $&$-68.08282 $&$-92.3323  $&$-118.778	 $\\
8     &$	  $&$	      $&$	  $&$-63.7371 $&$-73.2021 $&$-87.5117  $&$-117.5887 $&$-151.001	 $\\ 
9     &$	  $&$	      $&$	  $&$-80.8912 $&$-92.4162 $&$-109.4449 $&$-145.5587 $&$-186.4023 $\\ 
10    &$	  $&$	      $&$	  $&$-100.0783$&$-113.8625$&$-133.92080$&$-176.2966 $&$-224.937	 $\\
\end{tabular}
\label{tab:2senergy}
\end{table}

\newpage
\begin{table}
\caption{Total energies (a.u.) of the high-field ground states $\left|0_N\right>$ of positive ions $\rm A^+$ in strong magnetic fields.}
\begin{tabular}{@{}rlllllllll}
$Z   $&$\gamma=0.5$&$\gamma=1  $&$\gamma=2  $&$\gamma=5   $&$\gamma=10  $&$\gamma=20  $&$\gamma=50  $&\\ \noalign{\hrule}
2     &$-2.2346282$&$-2.4409898$&$-2.7888422$&$-3.5438677 $&$-4.3901481 $&$-5.5215956 $&$-7.5463093 $&\\
3     &$-5.640062 $&$-6.114623 $&$-6.894080 $&$-8.629427  $&$-10.651315 $&$-13.4297434$&$-18.525475 $&\\
4     &$-10.51258 $&$-11.31312 $&$-12.59206 $&$-15.42817  $&$-18.820184 $&$-23.612005 $&$-32.61959  $&\\
5     &$-16.9017  $&$-18.07243 $&$-19.93091 $&$-24.01520  $&$-28.93504  $&$-36.02020  $&$-49.63544  $&\\ 
6     &$-24.8433  $&$-26.4227  $&$-28.9235  $&$-34.40433  $&$-41.01061  $&$-50.62785  $&$-69.44195  $&\\
7     &$-34.3550  $&$-36.3826  $&$-39.5839  $&$-46.5957   $&$-55.04672  $&$-67.41737  $&$-91.94699  $&\\
8     &$	  $&$-47.9633  $&$-51.9215  $&$-60.5880   $&$-71.0369   $&$-86.37441  $&$-117.08457 $&\\ 
9     &$	  $&$	       $&$-65.9423  $&$-76.3802   $&$-88.9723   $&$-107.4850  $&$-144.8061  $&\\ 
10    &$	  $&$	       $&$-81.6509  $&$-93.9710   $&$-108.8443  $&$-130.7348  $&$-175.0743  $&\\
\noalign{\hrule}  
\noalign{\hrule}  
\noalign{\hrule}  
\noalign{\hrule}  
$Z   $&$\gamma=100$&$\gamma=200$&$\gamma=500 $&$\gamma=1000$&$\gamma=2000$&$\gamma=5000$&$\gamma=10000$&\\ \noalign{\hrule}  
2     &$-9.5605466$&$-12.071443$&$-16.2898727$&$-20.2706955$&$-25.028351 $&$-32.65713  $&$-39.548989$\\
3     &$-23.699944$&$-30.260769$&$-41.50393  $&$-52.323018 $&$-65.47657  $&$-86.9940   $&$-106.8134 $\\
4     &$-41.93414 $&$-53.90638 $&$-74.73619  $&$-95.07513  $&$-120.11947 $&$-161.7052  $&$-200.5709 $\\
5     &$-63.947265$&$-82.55711 $&$-115.33672 $&$-147.71743 $&$-187.99221 $&$-255.6619  $&$-319.6394 $\\ 
6     &$-89.51120 $&$-115.87500$&$-162.80039 $&$-209.6030  $&$-268.2990  $&$-367.8817  $&$-462.931  $\\
7     &$-118.45429$&$-153.5960 $&$-216.7194  $&$-280.1976  $&$-360.3670  $&$-497.502   $&$-629.454  $\\
8     &$-150.6447 $&$-195.5087 $&$-276.7565  $&$-359.0516  $&$-463.6191  $&$-643.7651  $&$-818.311  $\\ 
9     &$-185.9795 $&$-241.4411 $&$-342.6284  $&$-445.780   $&$-577.553   $&$-805.9918  $&$-1028.687 $\\ 
10    &$-224.3773 $&$-291.251  $&$-414.09358 $&$-540.0501  $&$-701.7295  $&$-983.5779  $&$-1259.8444$\\
\end{tabular}
\label{tab:Apground}
\end{table}

\begin{table}
\caption{Total energies (a.u.) of positive ions  $\rm A^+$ in strong magnetic fields 
in the fully spin polarized states $\left|2p_0\right>$.}
\begin{tabular}{@{}rlllllllll}
$Z   $&$\gamma=0.5 $&$\gamma=1   $&$\gamma=2   $&$\gamma=5  $&$\gamma=10  $&$\gamma=20  $&$\gamma=50  $&$\gamma=100 $&$\gamma=200 $\\ \noalign{\hrule}
3     &$-5.450607  $&$-5.790277	 $&$-6.354440  $&$-7.6155748$&$-9.07498561$&$-11.052577	$&$-14.60723  $&$-18.1514   $&$-22.5884 $\\
4     &$-10.71847  $&$-11.39964	 $&$-12.50590  $&$-14.969431$&$-17.896267 $&$-21.986880	$&$-29.579033 $&$-37.35540  $&$-47.2987	$\\
5     &$-17.58187  $&$-18.62668	 $&$-20.31984  $&$-24.06890 $&$-28.57270  $&$-35.01093	$&$-47.26504  $&$-60.06278  $&$-76.6646	$\\
6     &$-26.2094   $&$-27.6300	 $&$-29.9454   $&$-35.09561 $&$-41.30920  $&$-50.308209	$&$-67.773425 $&$-86.30316  $&$-110.6170$\\
7     &$-36.6424   $&$-38.4731	 $&$-41.4488   $&$-48.10689 $&$-56.17490  $&$-67.94617	$&$-91.12203  $&$-116.03024 $&$-149.0175$\\
8     &$	   $&$		 $&$-54.8620   $&$-63.1295  $&$-73.19399  $&$-87.94947	$&$-117.30577 $&$-149.19626 $&$-191.75146$\\ 
9     &$	   $&$		 $&$	       $&$-80.1774  $&$-92.3729	  $&$-110.32119	$&$-146.3062  $&$-185.7521  $&$-238.7295$\\ 
10    &$	   $&$		 $&$	       $&$-99.2581  $&$-113.7112  $&$-135.0539	$&$-178.0971  $&$-225.6432  $&$-289.8700$\\
\end{tabular}
\label{tab:2p0Penergy}
\end{table}

\begin{table}
\caption{Total energies (a.u.) of positive ions $\rm A^+$  in strong magnetic fields 
in the states $\left|1s^2\right>$.}
\begin{tabular}{@{}rllllllllllll}
$Z   $&$\gamma=0.5$&$\gamma=1 $&$\gamma=2 $&$\gamma=5 $&$\gamma=10$&$\gamma=20 $&$\gamma=50 $&$\gamma=100$&$\gamma=200$&\\ \noalign{\hrule}
3     &$-7.217983 $&$-7.164014$&$-6.962999$&$-5.850510$&$-3.110916$&$+3.748961 $&$+27.964647$&$+72.09337 $&$+164.66867$&\\
4     &$-14.49163 $&$-14.70591$&$-14.95181$&$-14.96820$&$-13.75773$&$-9.217910 $&$+10.42836 $&$+49.70820 $&$+135.95916$&\\
5     &$-24.2429  $&$-24.78674$&$-25.54108$&$-26.71999$&$-27.08417$&$-25.06410 $&$-10.65835 $&$+22.86883 $&$+101.46468$&\\ 
6     &$-36.5110  $&$-37.4273 $&$-38.7685 $&$-41.23663$&$-43.25931$&$-43.91255 $&$-35.28670 $&$-8.30098	 $&$+61.43184 $&\\
7     &$-51.3324  $&$-52.6586 $&$-54.6467 $&$-58.5397 $&$-62.33918$&$-65.81095 $&$-63.40519 $&$-43.64463 $&$+16.13450 $&\\
8     &$-68.725	  $&$-70.500  $&$-73.1912 $&$-78.6347 $&$-84.3435 $&$-90.78611 $&$-94.97381 $&$-83.03149 $&$-34.19097 $&\\ 
9     &$	  $&$-90.962  $&$-94.412  $&$-101.5242$&$-109.2779$&$-118.8537 $&$-129.9684 $&$-126.3622 $&$-89.3511  $&\\ 
10    &$	  $&$-114.054 $&$-118.316 $&$-127.2092$&$-137.1413$&$-150.0207 $&$-168.3763 $&$-173.5637 $&$-149.190  $&\\
\end{tabular}
\label{tab:s2Penergy}
\end{table}

\begin{table}
\caption{Total energies (a.u.) of the positive ions ${\rm A^+}$ in strong magnetic fields 
in the states $\left|2s\right>$.}
\begin{tabular}{@{}rllllllllllll}
$Z   $&$\gamma=0.5$&$\gamma=1 $&$\gamma=2 $&$\gamma=5 $&$\gamma=10$&$\gamma=20 $&$\gamma=50 $&$\gamma=100$\\ \noalign{\hrule}
3     &$-5.482414 $&$-5.725577$&$-6.125201$&$-7.168333$&$-8.489089$&$-10.34858 $&$-13.7869  $&$-17.2792	 $\\
4     &$-10.88665 $&$-11.48854$&$-12.39822$&$-14.49715$&$-17.22586$&$-21.18159 $&$-28.68072 $&$-36.4493	 $\\
5     &$-17.83551 $&$-18.82816$&$-20.35231$&$-23.60163$&$-27.76915$&$-34.02228 $&$-46.19941 $&$-59.0447	 $\\ 
6     &$	  $&$-27.9171 $&$-30.0961 $&$-34.69799$&$-40.35174$&$-49.03420 $&$-66.40440 $&$-85.0457	 $\\
7     &$	  $&$	      $&$	  $&$-47.83068$&$-55.11866$&$-66.30856 $&$-89.28664 $&$-114.3624 $\\
8     &$	  $&$	      $&$	  $&$-62.9964 $&$-72.1475 $&$-85.93862 $&$-114.84057$&$-146.90998$\\ 
9     &$	  $&$	      $&$	  $&$-80.1907 $&$-91.4315 $&$-108.01546$&$-143.08156$&$-182.6254 $\\ 
10    &$	  $&$	      $&$	  $&$-99.4125 $&$-112.9303$&$-132.5918 $&$-174.0451 $&$-221.4723 $\\
\end{tabular}
\label{tab:2spenergy}
\end{table}

\newpage

\begin{table}
\caption{Total energies (a.u.) of the neutral atoms and ions ${\rm A^+}$ 
at the crossover points of the ground state configurations.}
\begin{tabular}{@{}lllllllll}
$Z$&$\gamma$&Atomic state(s)&$-E{\rm (Atomic)}$&Ionic state(s)&$-E{\rm (A^+)}$\\ \noalign{\hrule}
2&0.711&$\left|0_N\right>$, $\left|1s^2\right>$&2.76940&$\left|0_N\right>$&2.32488\\
\noalign{\hrule}
3&2.153&$\left|0_N\right>$, $\left|1s^2\right>$&7.64785&$\left|0_N\right>$&7.00057\\
&2.0718&$\left|1s^2\right>$&7.65600&$\left|0_N\right>$, $\left|1s^2\right>$&6.94440\\
\noalign{\hrule}
4&4.567&$\left|0_N\right>$, $\left|1s^2\right>$&15.9166&$\left|0_N\right>$&15.07309\\
&4.501&$\left|1s^2\right>$&15.91625&$\left|0_N\right>$, $\left|1s^2\right>$&15.01775\\
\noalign{\hrule}
5&8.0251&$\left|0_N\right>$, $\left|1s^2\right>$&28.18667&$\left|0_N\right>$&27.16436\\
& 7.957&$\left|1s^2\right>$&28.17996&$\left|0_N\right>$, $\left|1s^2\right>$&27.10004\\
\noalign{\hrule}
6&18.664&$\left|0_N\right>$, $\left|2p_0\right>$&50.9257&$\left|0_N\right>$&49.50893\\
&14.536&$\left|2p_0\right>$&47.23836&$\left|0_N\right>$, $\left|2p_0\right>$&45.77150\\
&12.351&$\left|2p_0\right>$&45.07386&$\left|2p_0\right>$, $\left|1s^2\right>$&43.72095\\
&12.216&$\left|2p_0\right>$, $\left|1s^2\right>$&44.9341&$\left|1s^2\right>$&43.70075\\
\noalign{\hrule}
7&36.849&$\left|0_N\right>$, $\left|2p_0\right>$&84.4186&$\left|0_N\right>$&82.58182\\
&30.509&$\left|2p_0\right>$&79.34493&$\left|0_N\right>$, $\left|2p_0\right>$&77.41246\\
&17.429&$\left|2p_0\right>$&66.72786&$\left|2p_0\right>$, $\left|1s^2\right>$&65.26170\\
&17.398&$\left|2p_0\right>$, $\left|1s^2 2p_0\right>$&66.69306&$\left|1s^2\right>$&65.25362\\
\noalign{\hrule}
8&64.720&$\left|0_N\right>$, $\left|2p_0\right>$&130.6806&$\left|0_N\right>$&128.4054\\
&55.747&$\left|2p_0\right>$&124.1125&$\left|0_N\right>$, $\left|2p_0\right>$&121.69825\\
&23.985&$\left|2p_0\right>$, $\left|1s^2 2p_0\right>$&94.3773&$\left|2p_0\right>$&92.78308\\
&23.849&$\left|1s^2 2p_0\right>$&94.3336&$\left|2p_0\right>$, $\left|1s^2 2p_0\right>$&92.62502\\
\noalign{\hrule}
9&104.650&$\left|0_N\right>$, $\left|2p_0\right>$&191.8770&$\left|0_N\right>$&189.1446\\
&92.624&$\left|2p_0\right>$&183.6944&$\left|0_N\right>$, $\left|2p_0\right>$&180.7819\\
&31.735&$\left|2p_0\right>$, $\left|1s^2 2p_0\right>$&128.1605&$\left|2p_0\right>$&126.4414\\
&31.612&$\left|1s^2 2p_0\right>$&128.1125&$\left|2p_0\right>$, $\left|1s^2 2p_0\right>$&126.2897\\
\noalign{\hrule}
10&159.138&$\left|0_N\right>$, $\left|2p_0\right>$&270.220&$\left|0_N\right>$&267.0112\\
&143.604&$\left|2p_0\right>$&260.2740&$\left|0_N\right>$, $\left|2p_0\right>$&256.8459\\
&40.672&$\left|2p_0\right>$, $\left|1s^2 2p_0\right>$&168.4734&$\left|2p_0\right>$&166.6327\\
&40.559&$\left|1s^2 2p_0\right>$&168.4217&$\left|2p_0\right>$, $\left|1s^2 2p_0\right>$&166.4863\\
\end{tabular}
\label{tab:trans}
\end{table}

\begin{table}
\caption{Absolute values of the 
total energies (keV) of the high-field ground states of neutral atoms in strong magnetic fields 
compared with the literature. $B_{12}=B/(10^{12}{\rm G})$.}
\begin{tabular}{@{}rlllllllllll}
&\multicolumn{3}{c}{$B_{12}=0.1$}&\multicolumn{2}{c}{$B_{12}=0.5 $}&\multicolumn{2}{c}{$B_{12}=1   $}&\multicolumn{2}{c}{$B_{12}=2.3505\ (\gamma=1000)$}&\multicolumn{2}{c}{$B_{12}=5   $}\\  
\cline{2-4} \cline{5-6} \cline{7-8} \cline{9-10} \cline{11-12} 
\multicolumn{1}{c}{$Z$}&
\multicolumn{1}{c}{\ IS ($\left|2p_0\right>$)}&
\multicolumn{1}{c}{IS}&
\multicolumn{1}{c}{NKL}&
\multicolumn{1}{c}{IS}&
\multicolumn{1}{c}{NKL}&
\multicolumn{1}{c}{IS}&
\multicolumn{1}{c}{NKL}&
\multicolumn{1}{c}{IS}&
\multicolumn{1}{l}{DHG}&
\multicolumn{1}{c}{IS}&
\multicolumn{1}{c}{NKL}\\ \noalign{\hrule}  
1     &             &0.07781&0.0761\ \ \ &0.13114&0.130\ \ \ &0.16222&0.161\ \ \ &0.20851&0.206\ \ \ &0.25750&0.2550\\
2     &             &0.26387&0.255 &0.46063&0.454&0.57999&0.574&0.76279&0.754&0.96191&0.9580\\
3     &             &0.54042&0.516 &0.96180&0.944&1.22443&1.209&1.63429&1.611&2.08931&2.0760\\
4     &             &0.89833&0.846 &1.61624&1.580&2.07309&2.042&2.79610&2.746&3.61033&3.5840\\
5     &             &1.33229&1.238 &2.41101&2.347&3.10924&3.054&4.22674&4.139&5.49950&5.4560\\ 
6     &             &1.83895&1.678 &3.33639&3.22 &4.31991&4.20 &5.90872&5.773&7.73528&7.60\\
7     &             &2.41607&2.17  &4.38483&4.22 &5.69465&5.54 &7.82757&     &10.29919&10.20\\
8     &\ \ \ 3.08253&3.06214&2.71  &5.55032&5.32 &7.22492&7.02 &9.97107&     &13.17543&13.00\\ 
9     &\ \ \ 3.82966&3.77607&3.36  &6.82794&6.51 &8.90360&8.63 &12.32880&    &16.34997&16.10\\ 
10    &\ \ \ 4.65087&4.55698&      &8.21365&7.819&10.72452&10.39&14.89168&   &19.81072&19.57\\
\end{tabular}
{\footnotesize
IS -- present work\\
NKL -- results by Neuhauser, Koonin and Langanke \cite{Neuhauser}\\
DHG -- results by Demeur, Heenen and Godefroid \cite{Godefroid}\\
$\left|2p_0\right>$ -- results for states $\left|2p_0\right>$ 
at the points where they are the ground states
}
\label{tab:Neucomp}
\end{table}

\end{document}